\documentclass[twocolumn,superscriptaddress,showpacs,prl,amsmath,amssymb,wasysymb]{revtex4}
\usepackage{graphicx}
\usepackage{bm}
\begin{document}

\title{
Infinite family of persistence exponents for interface fluctuations}

\author{ M. Constantin}
\affiliation{
Condensed Matter Theory Center, 
Department of Physics, University of Maryland, College Park, Maryland 
20742-4111}
\affiliation{Materials Research Science and Engineering Center, 
Department of Physics, University of Maryland, College Park, MD 20742-4111
}
\author{ S. \surname{Das Sarma}}
\affiliation{
Condensed Matter Theory Center, 
Department of Physics, University of Maryland, College Park, Maryland 
20742-4111}
\author{ C. Dasgupta}
\altaffiliation{Permanent address: Department of Physics, Indian 
Institute of Science, Bangalore 560012, India
}
\affiliation{
Condensed Matter Theory Center, 
Department of Physics, University of Maryland, College Park, Maryland 
20742-4111}
\author{ O. Bondarchuk}
\affiliation{Materials Research Science and Engineering Center, 
Department of Physics, University of Maryland, College Park, MD 20742-4111
}
\author{ D. B. Dougherty}
\affiliation{Materials Research Science and Engineering Center, 
Department of Physics, University of Maryland, College Park, MD 20742-4111
}
\author{ E. D. Williams}
\affiliation{Materials Research Science and Engineering Center, 
Department of Physics, University of Maryland, College Park, MD 20742-4111
}

\begin{abstract}
We show experimentally and theoretically that the persistence of 
large deviations in equilibrium step fluctuations is characterized 
by an infinite family of independent exponents. These exponents are obtained by
carefully analyzing dynamical experimental images of Al/Si(111) 
and Ag(111) equilibrium steps fluctuating at high (970K) and 
low (320K) temperatures respectively, and by quantitatively 
interpreting our observations on the basis of the corresponding 
coarse-grained discrete and continuum theoretical models for 
thermal surface step fluctuations under attachment/detachment 
(``high-temperature'') and edge-diffusion limited kinetics 
(``low-temperature'') respectively. 
\end{abstract}


\maketitle

Persistence is a fundamental and powerful concept in the 
stochastic dynamics of non-Markovian statistical processes 
\cite{1}. Recently, the persistence concept has been applied to statistical 
studies of spatially extended non-equilibrium systems both theoretically \cite{3} 
and experimentally \cite{4}. Loosely speaking, persistence is the probability 
$P(t)$ that a stochastic variable, which is found to have 
a ``positive'' value at time $t_{0}$, will stay ``positive'' 
throughout a time interval $t$ up to time 
$t_{0} + t$. In most steady-state (``stationary'') 
situations, the persistence probability is found to decay 
in time in a power-law manner, $P(t) \sim t^{-\theta}$, for large $t$, 
where the {\it persistence exponent} $\theta$ 
is a highly nontrivial exponent characterizing the relevant 
stochastic dynamics under consideration. 
The concept of persistence has been used \cite{6,7,8} to study 
the interesting problem of thermally fluctuating interfaces 
where steps on vicinal surfaces undergo random thermal 
motion in equilibrium \cite{5}. The step persistence probability is 
the probability $P(t_{0},t_{0}+t)$ that a given lateral step position 
$x$ with a height (i.e. step fluctuation measured from 
the equilibrium step position) $h(x,t_{0})$ 
at time $t_{0}$ does not return to this value up to a 
later time $t_{0} + t$. With no loss of generality we 
will set $t_{0}=0$ from now on, assuming that thermal 
equilibrium has been achieved in the step fluctuations 
and we are discussing steady-state stationary properties. 
The resulting persistence probability $P(t) \equiv P(0,t)$ has 
recently been studied experimentally using dynamical scanning 
tunneling microscopy (STM) in two systems: Al steps on Si $(111)$ 
surface at high temperatures $\sim 970$K \cite{7} and Ag $(111)$ surface 
at low temperatures $\sim 320$K \cite{8}. In the first case, the 
step fluctuations dominated by atomistic attachment and detachment 
(AD) at the step edge are known \cite{5} to be well described by the 
coarse-grained second-order non-conserved linear Langevin equation 
\begin{equation}
\label{eq1}
\frac {\partial h(x, t)} {\partial t} = \frac{\partial^{2} h(x,t)}
{\partial x^2} + \eta(x, t),
\end{equation}
\noindent where $\eta(x,t)$ with
$\langle \eta(x, t) \eta(x^{'},t^{'}) \rangle \propto \delta(x-x^{'}) 
\delta(t-t^{'})$ is the usual uncorrelated random Gaussian 
noise corresponding to the non-conserved white noise associated 
with the random AD process. Low temperature step fluctuations 
dominated by the step edge diffusion (ED) mechanism are, on 
the other hand, described by a fourth order conserved 
linear Langevin equation:
\begin{equation}
\label{eq2}
\frac {\partial h(x, t)} {\partial t} = -\frac{\partial^{4} h(x,t)}
{\partial x^4} + \eta_{c}(x, t),
\end{equation}
\noindent where $\eta_c(x,t)$ with 
$\langle \eta_{c}(x,t) \eta_{c}(x^{'},t^{'}) \rangle \propto 
\nabla^{2} \delta(x-x^{'}) \delta(t-t^{'})$ is a conserved 
noise associated with atomic diffusion along the step edge.
From a quantitative analysis of the digitized STM step images as 
a function of time, the persistence exponent 
was found to be $\theta = 0.77 \pm 0.03$ \cite{7} and 
$\theta = 0.87 \pm 0.02$ \cite{8}, respectively, for the 
high-temperature AD mechanism and low-temperature ED mechanism. 
These measured step fluctuation persistence exponents agree reasonably well with those 
found \cite{6} from kinetic Monte Carlo simulations of the corresponding 
discrete solid-on-solid models: $\theta \simeq 0.75$ 
(for Eq.~(\ref{eq1})) and $\theta \simeq 0.86$ (for Eq.~(\ref{eq2})).
These results are in agreement with
a postulated (and numerically verified) relation \cite{6} between 
persistence in the steady state and dynamic scaling \cite{9} in the 
pre-stationary transient regime. It is believed \cite{6} that, 
at least for linear Langevin equations, 
the persistence exponent $\theta$ is equal to $(1-\beta)$ where $\beta$
is the exponent that describes the initial power-law growth of the interface
width \cite{9} in the transient regime
($\beta=1/4$ for Eq.~(\ref{eq1}) and $\beta=1/8$ for Eq.~(\ref{eq2})).

This striking agreement between experimentally obtained and theoretically
predicted values of the persistence exponent demonstrates the overall 
excellent consistency among theory, experiment, and simulations 
in this problem, but also brings up a key question 
regarding persistence studies \cite{6,7,8} of surface 
fluctuations: Is persistence really an independent 
(and new) conceptual tool in studying surface fluctuations, 
or, is it just an equivalent (perhaps even complementary) 
way of studying dynamic scaling \cite{5,9} of height correlations? 
In this Letter we present new theoretical and experimental persistence 
results on surface step fluctuations that fundamentally transcend 
any dynamic scaling considerations, establishing in the process the 
existence of a novel and nontrivial infinite family (i.e. a continuous set) 
of persistence exponents for equilibrium step fluctuations. We carry out 
quantitative analyses of (digitized) dynamical STM images of step fluctuations 
both for high-temperature (Al/Si) and low-temperature 
(Ag) equilibrium surfaces, and compare in details the 
experimental results with those we have obtained from numerical 
integration of the corresponding Langevin equations and discrete 
stochastic Monte Carlo simulations of corresponding atomistic cellular automata 
type models in the same dynamical universality classes \cite{5,9}. 
All three sets of persistence results 
agree very well for both high and low temperature 
equilibrium step fluctuations, establishing persistence 
(particularly, the infinite family of persistence exponents) as a 
potentially powerful tool (rivaling, perhaps even exceeding, 
in utility the well-studied dynamic scaling approach) in studying 
dynamical interface fluctuation processes. 

The infinite family of persistence exponents we study 
here is based on the concept of {\it persistence of large 
deviations} introduced recently by Dornic and Godreche \cite{10} 
in the context of kinetic Glauber-Ising dynamics of 
magnetization coarsening (a closely related idea, 
that of sign-time distribution, was developed in Ref.~\cite{12}). 
For equilibrium step fluctuations, we define the probability of 
persistent large deviations, $P(t,s)$, as the probability 
for the ``average sign'' $S_{\text{av}}$ of the height fluctuation 
to remain above a certain pre-assigned value ``$s$'' up to time $t$:
\begin{equation}
\label{eq4}
P(t,s) \equiv \hbox{Prob}~\lbrace ~S_{\text{av}}(t^{\prime}) \geq s,
~\forall t^{\prime} \leq t~ \rbrace,
\end{equation}
\noindent where 
\begin{equation}
\label{eq5}
S_{\text{av}}(t) \equiv t^{-1} \int_{0}^{t} {dt^{\prime}~S(t^{\prime})},
\end{equation}
\noindent and
\begin{equation}
\label{eq6}
S(t) \equiv \hbox{sign}~[h(x,t_{0} + t) - h(x,t_{0})] ,
\end{equation}
\noindent where with no loss of generality, we set $t_{0} = 0$. 
Although the dynamical variable $S(t)$ above is defined for a 
particular lateral position $x$, 
we take a statistical ensemble average over all lateral positions 
to obtain a purely time dependent dynamical quantity $P(t,s)$.
Since $S_{\text{av}}(t) \in [-1,1]$, the probability $P(t,s)$ is 
defined for $-1 \le s \le 1$. For $s=1$ we recover our earlier simple 
definition of persistence used in Refs.~\cite{6,7,8}: 
$P(t) \equiv P(t,s=1)$ measures the probability of the height 
fluctuation remaining above zero (``positive'') throughout 
the whole time interval. For $s=-1$ the probability $P(t,s=-1)$ 
is trivially equal to unity for all $t$. The generalized 
probability function, $P(t,s)$, defined in 
Eqs.~(\ref{eq4})--(\ref{eq6}) above, leads to a continuous 
family or hierarchy of persistent {\it large} deviations exponents, $\theta_{l}(s)$, 
{\it provided} the steady-state decay of $P(t,s)$ in time follows a 
power law, $P(t,s) \sim t^{-\theta_{l}(s)}$. As we show 
below, this indeed happens for equilibrium step fluctuation phenomena, 
allowing us to define and measure the non-trivial persistence exponent 
$\theta_{l}(s),\,-1 \le s \le 1$, that varies continuously 
between $\theta_{l}(s=-1) \equiv 0$ and $\theta_{l}(s=+1) \equiv \theta$ 
where $\theta$ is the usual persistence exponent. Clearly, $P(t,s)$ and 
$\theta_{l}(s)$ are natural generalizations of the 
persistence probability $P(t)$ and the persistence exponent 
$\theta$, respectively, to the broader concept of distribution 
of residence times with limiting behavior (i.e. $s=1$) 
determining the usual persistence exponent.

\begin{figure}
\includegraphics[height=16.5cm,width=7.8cm]{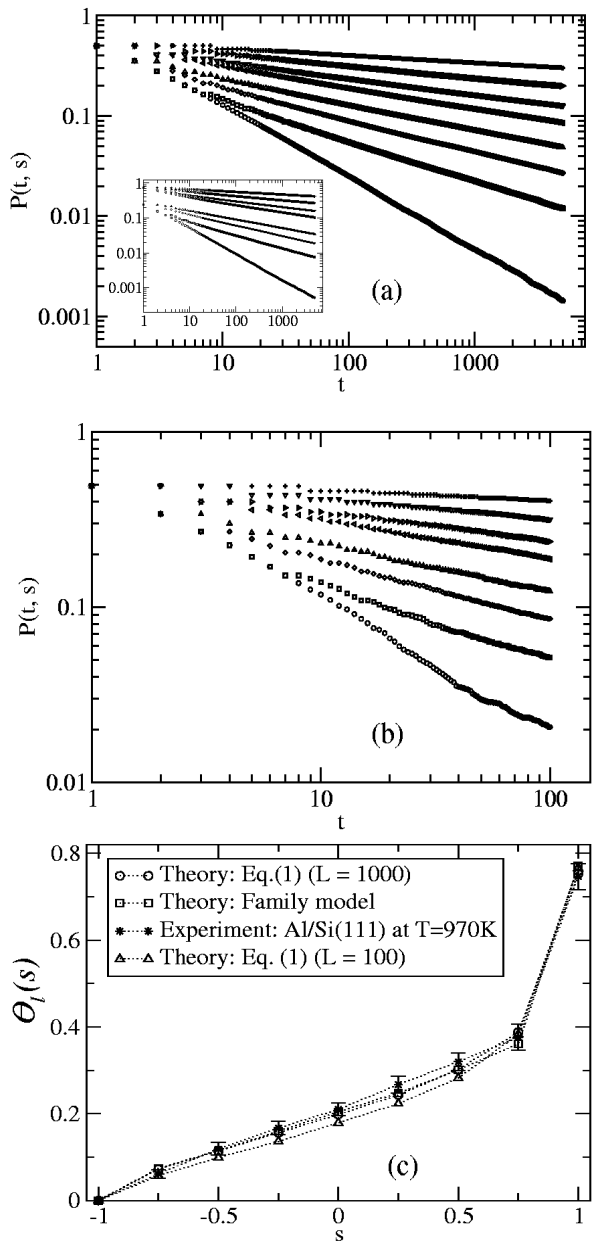}
\caption{\label{fig1} Log-log plots of the persistent large deviations 
probability, $P(t,s)$, for high-temperature surface step fluctuations via the
AD mechanism, shown as a function of time ($t$) for different values 
of the average sign constraint parameter 
$s=1.0,~0.75,~0.5,~0.25,~0,-0.25,-0.5,-0.75$ (from the bottom 
to the top). (a) The continuum Langevin equation of Eq.(\ref{eq1}) 
(main figure) and the stochastic discrete Family model (inset); 
(b) experimental data from digitized dynamical STM 
step images of Al on Si(111) surface; and (c) comparison 
of the various sets of results for $\theta_l$ as a function
of $s$. The error bars shown for the 
experimental data are obtained from variations of the local 
slope of the $\log P(t,s)$ vs $\log t$ plots. Simulation results 
for two sample sizes ($L=100$ and $L=1000$) are shown to illustrate 
that the use of small samples leads to an underestimation of $\theta_l(s)$.}
\end{figure}

\begin{figure}
\includegraphics[height=16cm,width=7.5cm]{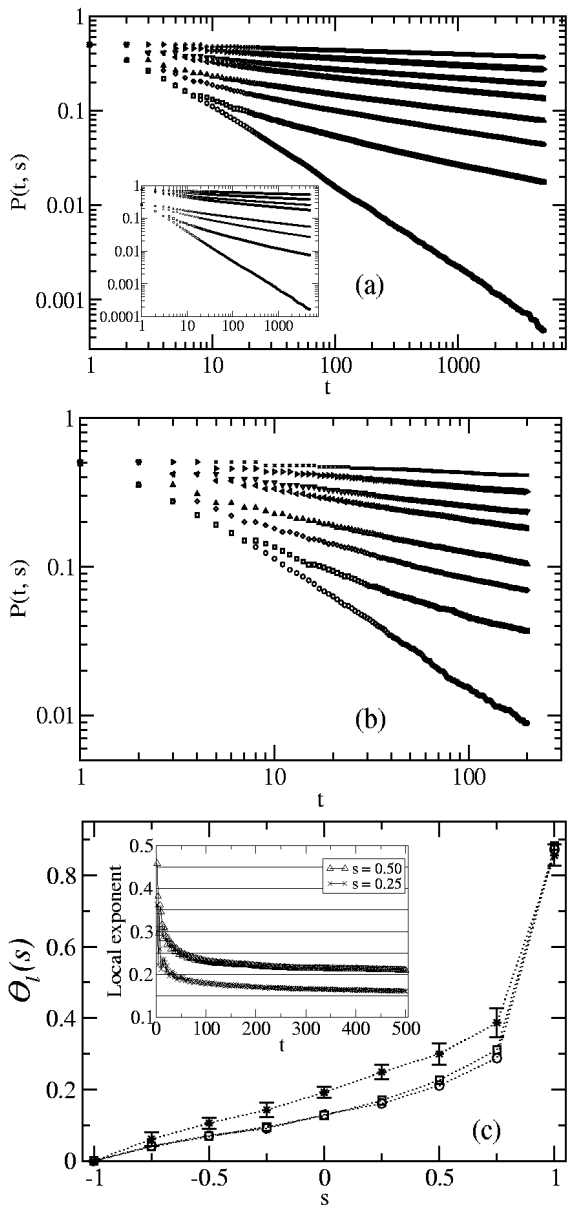}
\caption{\label{fig3}
The same as in Fig.~\ref{fig1} for the low-temperature
step fluctuations via the ED mechanism. (a) The continuum Langevin 
equation of Eq.~(\ref{eq2}) (main figure) 
and the discrete stochastic Racz model (inset); (b) experimental data for the
Ag(111) step fluctuations; and (c) comparison of the
three sets of results for $\theta_l$: experiment (star), 
stochastic Racz model (square) and Eq.~(\ref{eq2}) 
(circle). The inset shows the time-dependence of the 
{\it local} exponent $\theta_l(s,t)$ obtained from simulations 
of the Langevin equation for $s=0.25$ and $s=0.50$.}
\end{figure}

In this Letter, we have carried out the first application of the 
persistent large deviations concept to the equilibrium step 
fluctuations phenomenon.  We also report the first experimental 
measurements of $P(t,s)$ and $\theta_{l}$ for {\it any} stochastic 
system. Our results for the two distinct types 
of step fluctuation processes (the high-temperature 
AD and low-temperature ED situations, described \cite{5} by 
Eqs.~(\ref{eq1}) and (\ref{eq2}), respectively) are shown in 
Figs.~\ref{fig1} and \ref{fig3}. The details of the experimental procedure 
for extracting the persistence probability from the digitized 
dynamical STM images are described in Refs.~\cite{7,8} for the 
same two systems, and are therefore not repeated here. We only 
mention that in order to obtain the probability of persistent large 
deviations, we had to use a substantially larger number ($\sim 100$)  
of STM data sets than that used in Refs.~\cite{7,8}. As mentioned 
earlier, we compare the experimental results with two theoretical models 
in each case: The continuous Langevin equation (Eq.~(\ref{eq1}) or (\ref{eq2})) 
and a discrete stochastic model which is theoretically known 
\cite{6,7,8,5,9} to belong to the same universality class 
(in the dynamic scaling sense) as the continuum equation. 
For the high-temperature AD step fluctuations, described 
on a coarse-grained scale by Eq.~(\ref{eq1}) (sometimes referred 
to as the Edwards--Wilkinson equation \cite{9}), the discrete 
stochastic model we use is the extensively studied 
solid-on-solid Family model \cite{9,13}. 
For the low-temperature ED case, described on a coarse-grained scale 
by Eq.~(\ref{eq2}) (sometimes referred to as the 
Mullins--Herring equation with conserved noise \cite{9}), 
the discrete stochastic model we use is the well-studied Racz model \cite{14}. 
Our results are obtained from numerical integration of the 
Langevin equations using the simple Euler scheme \cite{6}, 
and standard Monte Carlo simulation \cite{6} of the atomistic 
models. Typical system sizes used in the numerical work are $\sim 1000$ 
for Eq.(\ref{eq1}) and $\sim 100$ for Eq.(\ref{eq2}), and the number of 
independent runs used in the calculation of persistence probabilities is 
$\sim 1000$.

In Fig.~\ref{fig1} we show in the top two panels our measured 
(panel (b)) and calculated (panel (a), main figure and inset) 
persistent large deviation probability $P(t,s)$ as a function of 
time $t$ for the high-temperature step fluctuations case. Each panel 
shows eight different log-log plots of $P(t,s)$ against $t$ for eight 
different values of the average sign parameter 
$s~(=+1,+0.75,...,-0.75$ from the bottom to the top). As mentioned 
before, the case $s=+1$ (the bottom-most curve) corresponds to the usual
persistence probability $P(t)$,  and therefore the results shown in 
Fig.~\ref{fig1} for $s=1$ are already known. The results for all the 
other values of $s$ are new and non-trivial. The linearity of the 
log-log $P(t,s)$ vs. $t$ plots immediately implies that 
$P(t,s) \sim t^{-\theta_{l}(s)}$. The different sets of results for 
$\theta_l$ as a function of $s$ are shown in 
Fig.~\ref{fig1}(c). The excellent agreement among the various data sets shown 
in Fig.~\ref{fig1}(c) is the most important new quantitative 
result of our work. This means that the high-temperature 
step fluctuation phenomenon via the AD mechanism is indeed 
described by the Edwards--Wilkinson equation (and therefore 
also by the discrete Family model), not just in the sense 
of the dynamical universality class (as defined by specific 
exponent values, e.g. $\beta$ and $\theta$) but more 
importantly for the infinite family of persistent large 
deviations exponents as defined by the continuous 
{\it function} $\theta_{l}(s)$. This striking agreement between 
experiment and theory for a continuous family of exponents definitely 
establishes persistent large deviations studies as a new and effective 
tool for studying dynamical fluctuations of nanoscale systems. 

In Fig.~\ref{fig3} we present results similar 
to those in Fig.~\ref{fig1}, but now for the 
ED mechanism step fluctuation data along 
with the corresponding theoretical results for the continuum 
Langevin equation defined by Eq.~(\ref{eq2}) and the discrete 
stochastic Racz model which are known \cite{6,8,5,13} to be in 
the same dynamic universality class as the low-temperature 
step fluctuation process. The same description and explanation 
given above for Fig.~\ref{fig1} apply now to Fig.~\ref{fig3} where 
$\theta_{l}(s=1) \approx 0.875$, which agrees with 
recent experimental measurements \cite{8} of the usual 
persistence exponent of low-temperature step fluctuations 
on Ag and Pb surfaces. The experimental and theoretical results 
for the continuous function $\theta_{l}(s)$, shown in Fig.~\ref{fig3}(c), 
exhibit qualitative agreement, with the experimental exponent values 
for $s<1$ being slightly larger than the theoretical ones. 
There are several possible explanations for this difference between 
experimental and theoretical results. There are reasons to expect
that increasing the dynamic range of the experimental $P(t,s)$ beyond two 
decades in $t$ (this range is limited by noise problems 
inherent in dynamic STM imaging) would bring theory and experiment into 
closer agreement. To illustrate this possibility, we show in the inset 
of Fig.~\ref{fig3}(c) the time-dependence of the {\em local} 
exponent $\theta_l(s,t) = d \log[P(s,t)]/ d \log t$, obtained from
simulations of the Langevin equation for two values of $s$ for which the
difference between theory and experiment is large. The local exponent 
is found to {\em decrease} with time before reaching a constant value 
at large $t$. We have checked that the experimental data show similar 
behavior for all $s<1$, which implies that the effective exponent values 
obtained from power-law fits of relatively short-time data would be 
larger than the true long-time values. Indeed, we have found that 
fits of the simulation data over the range $20 \le t \le 200$ (this is
the range used in obtaining $\theta_l(s)$ from the experimental data) yield 
values of $\theta_l(s)$ that are higher and closer to the experimental 
values. A second possibility is that the smallness of the sample size used
in the simulations leads to an underestimation of the values 
of $\theta_l(s)$, as in Fig.~\ref{fig1}(c). Unfortunately, the impossibility
of equilibrating much larger samples of these models with very slow dynamics
prevents us from checking this explicitly. Another possibility that we can not
rule out is that the Ag(111) equilibrium step fluctuations do not precisely follow
the theoretical models of edge-diffusion limited kinetics. 
Further experimental and theoretical investigations would be needed for settling 
this issue. We should emphasize, however, that given the severe complexity 
in measuring any power law exponents associated with surface 
step fluctuation dynamics, the overall agreement between theory and 
experiment is quite good.

In summary, we have established the concept (and the usefulness) 
of an infinite family of persistent large deviations exponents 
for height fluctuations in equilibrium surface step dynamics 
phenomenon. The impressive agreement between theory and 
experiment indicates that the persistent large deviations 
probability (and the corresponding exponents) may very well 
be an extremely powerful tool in characterizing and 
understanding other stochastic fluctuation phenomena, e.g., 
kinetic surface roughening in nonequilibrium growth. 
In contrast to other dynamical approaches, the technique 
developed in this Letter leads to a continuous family of 
exponents (a continuous function rather than one or two isolated 
independent exponents (as in the dynamic scaling approach) and is 
therefore a much more stringent test of theoretical ideas, and 
also perhaps provides a deeper level of probing the dynamics of 
fluctuation problems. 

This work is partially supported by the NSF-DMR-MRSEC at 
the University of Maryland.

\end{document}